\documentclass[]{spie}  %>>> use for US letter paper
\usepackage{amsmath,amsfonts,amssymb}
\usepackage{graphicx}
\usepackage[colorlinks=true, allcolors=blue]{hyperref}

\title{The survey operation software system development for Prime Focus Spectrograph (PFS) on Subaru Telescope}

\author[a]{Atsushi Shimono}
\author[a]{Naoyuki Tamura}
\author[b]{Naruhisa Takato}
\author[a]{Naoki Yasuda}
\author[a]{Nao Suzuki}
\author[c]{Craig P. Loomis}
\author[c]{Robert H. Lupton}
\author[a]{Yuki Moritani}
\author[a]{Kiyoto Yabe}
\affil[a]{Kavli Institute for the Physics and Mathematics of the Universe (WPI), The University of Tokyo Institutes for Advanced Study, The University of Tokyo, 5-1-5 Kashiwanoha, Kashiwa, Chiba, 277-8583, Japan}
\affil[b]{Subaru Telescope, National Astronomical Observatory of Japan, 650 North A'ohoku Place, Hilo, HI 96720, United States}
\affil[c]{Princeton University, Department of Astrophysical Sciences, Peyton Hall, 4 Ivy Lane, Princeton, NJ 08544, United States}

\authorinfo{Further author information: (Send correspondence to A.S.)\\A.S.: E-mail: atsushi.shimono@ipmu.jp, Telephone: 81 4 7136 5976}

\begin{document} 
\maketitle

\begin{abstract}
The Prime Focus Spectrograph (PFS) is a wide-field, multi-object spectrograph accommodating 2394 fibers to observe the sky at the prime focus of the Subaru telescope. The software system to operate a spectroscopic survey is structured by the four packages: Instrument control software, exposure targeting software, data reduction pipeline, and survey planning and tracking software. In addition, we operate a database system where various information such as properties of target objects, instrument configurations, and observation conditions is stored and is organized via a standardized data model for future references to update survey plans and to scientific researches. In this article, we present an overview of the software system and describe the workflows that need to be performed in the PFS operation, with some highlights on the database that organizes various information from sub-processes in the survey operation, and on the process of fiber configuration from the software perspectives.

\end{abstract}


\section{INTRODUCTION}
\label{sec:intro}  % \label{} allows reference to this section

The Prime Focus Spectrograph (PFS \cite{2016SPIE990859}) is a wide field multi-object spectrograph with 2394 science fibers over a single hexagonal field of view (FoV) of 1.3 degree effective diameter, and is scheduled to be mounted at the prime focus of the 8.2m Subaru telescope. In parallel to the PFS will be operated as a facility instrument of the Subaru, a large survey program under the Subaru Strategic Program (SSP), of which the current framework is up to 5 years with $\sim$300 nights in total, is assumed to be operated using the PFS. 

The PFS instrument is consisted with mainly three hardware: the Prime Focus Instrument (PFI) 
to be mounted at the prime focus of the Subaru telescope, spectrograph system (SpS) which is 
four identical modules at so-called IR4 floor in a dome where spectrographs of FMOS is placed, 
and the metrology camera system (MCS) to be mounted at the Cassegrain focus. 

\begin{figure}
\begin{center}
\includegraphics{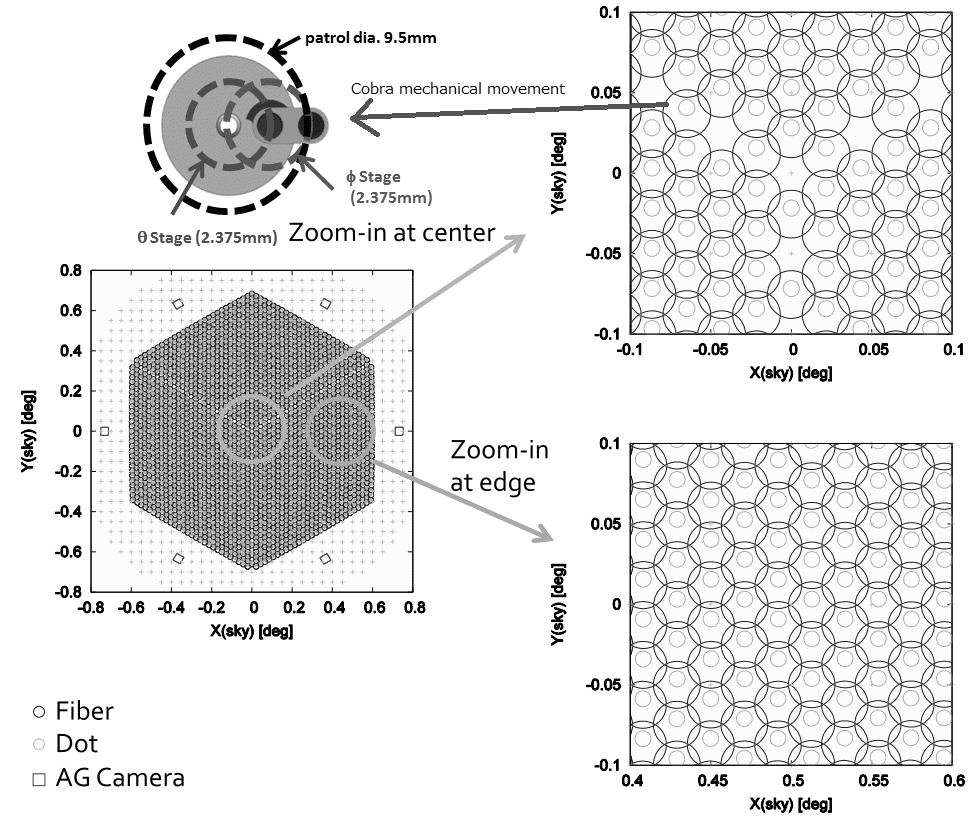}
% use 200dpi, in PNG header
\end{center}
\caption
{\label{fig:cobra}
Mapping of Cobra and its patrol area on the sky. 
}
\end{figure}

The PFI of PFS has 2394 "Cobra" positioners to move fibers to targets, which is consisted of 
two rotational stages to cover a circular patrol area per each Cobra \cite{2014SPIE.9151E..1YF} 
(Figure.~\ref{fig:cobra}.). These positions on the focal plane is fixed, assignment of targets 
to Cobras and assignments coordination over multiple exposures on each tile is a key for 
efficient survey operation. 

The PFS survey program \cite{2014PASJ...66R...1T} is mostly categorized into two modes on 
operational point of view: mostly uniform total exposure time and continuing till required 
signal to noise ratio with wide variety of total exposure time from less than an hour to 
10 hours. In any case, targets could be marked as done whenever their estimated signal to 
noise ratio satisfy required number, and target assignment for these Cobras need to be updated. 
To perform such frequent reassignment of targets, a database system to track progress on 
data acquisition and also additional intermediate information during target assignment is a 
key for efficient operation, like switching targets for upcoming exposure of each tile per 
object but not by entire configuration. If we keep tile definition, which could be defined 
by a center position of a tile and a position angle, a patrol area per each Cobra could be 
fixed at the sky coordinate, and we can list spare targets for each patrol area for ease of 
replace target objects of exposure configuration without switching an entire configuration 
to another. 

In the survey by PFS, a number of targets both observed at once and assigned to one exposure 
is significantly large, capabilities for on-site automatic rearrangements and redesignings, 
such that switching scientific targets marked as succeeded to acquire required signal to 
noise ratio, are important for efficient survey operation. For these, PFS is planning to 
have somehow simpler version or module picked up for works of the SPT and the DRP packages.

PFS will be a facility instrument at the Subaru Telescope observatory, so the software system has been under development with potential requirements from the general open-use framework envisioned as well as those dedicated to a long-term survey which the PFS collaboration is planning to conduct.  
The overall survey operation is coordinated by four internal software packages that are designed to be loosely coupled to each other. The definitions of these four packages have still been evolving in detail as the survey operation concept is updated and additional requirements need to be accommodated, but the main bodies have been consolidated as follows. 
The instrument control software will execute the observation communicating not only with the individual PFS hardware control subsystems but also with the sequencing system for the entire Subaru telescope system. 
The exposure targeting software provides assignments of the fibers to target objects at individual exposures taking into account various factors such as flexures of the telescope and the prime focus instrument of PFS, the residual effects of atmospheric dispersion after corrections by Atmospheric Dispersion Corrector (ADC), and so on. This software will aim to efficiently distribute the required fiber hours of target objects to multiple exposures, expecting that the telescope pointing center and subsequently the observing field of each exposure is given at a high-level layer (e.g. from observers and survey operators). 
The data reduction pipeline will reduce the spectral images from exposures, transform them into one-dimensional calibrated spectra and measure spectroscopic parameters. A subset of the functions in this pipeline will also be used for on-site data quality assessment and assurance on the individual exposure basis mainly using the data of relatively bright stars that are observed in the same exposures with faint science targets. 
The survey planning and tracking software is to manage survey plans according to the progresses and updates reconciling the inputs from observers and survey operators with the outputs from the data quality assessment and assurance.
%based on the field tiling definition supplied from science survey operators. 
%This package will aim to efficiently distribute the required fiber hours of target objects to multiple exposures, and also will track progress of the entire survey to update sequences and configurations of planned exposures.

These software packages are consisted of functional modules and sequences, some examples of which are taking a science or calibration exposure by the instrument control software, and mapping fibers to science targets by the exposure targeting software. These modules and sequences will actually be used by instrument operators from the observatory or observers themselves, depending probably on the program type (either a long-term survey or a PI-type short program), in such a way that the selected modules will be called to provide required functions from the entire packages both internally and externally. The interfaces between different modules are defined by a standardized data model so that the information will be managed on the target-by-target basis.
The modules of the instrument control software will be mostly connected to each other via a messaging hub server and will be called from sequences through it. For this hub server, we selected the tron system that has been operating at SDSS. Its sequencer is built in as a module connected to the hub server, and some modules from other PFS software packages will be called from this sequencer locally, without adding any wrappers. In the survey planning and tracking software and the data reduction pipeline, sequencers will call required modules and packages either locally or from a parallel computing system, without this hub server involved. 

To carry out survey operations efficiently, a database system will be operated to store necessary information such as the photometry of target objects and the instrument configuration at a given exposure and will be referred to by the software packages. This database system may be located physically at multiple places. In particular for observation execution, the database will need to be at the site of observation for secure operation against unexpected network interruptions from the outside of the observatory. Data flows and synchronization between these multiple databases will be studied and developed as the other software packages and data models are matured.

Entire PFS survey operation software is consisted with four software packages 
\cite{2012SPIE.8451E..3FS} : instrument control software (ICS) package, data reduction 
package (DRP), exposure targeting software (ETS) package, and survey planning and tracking 
(SPT) package. The SPT package is required to coordinate entire large survey programs, such 
by assigning target objects into multiple exposures, building exposure configurations with 
calibration and guiding targets by interfacing to the ETS, and feeding back to exposure 
configurations from data quality assurance by the DRP. The SPT package will be a collection 
of tools for survey processing and is designed based on one integrated database system, 
contains all information on the survey such as lists of target objects, tiling and exposure 
configurations, and properties from processed data. The ETS package works on a single field, 
which will be one of pre-defined tiles for survey or a defined field to observe for an 
individual observation program, to construct one full configuration of exposure consisted 
with science target assignments, calibration targets (e.g. sky, standard star, or calibration 
stars), and object for field acquisition and telescope guiding, from supplied list of 
science targets. Exposure configurations built by the ETS will be executed by the ICS on 
the site, which controls entire instrument with interfacing to a control software of each 
component under interaction with Subaru telescope control system (so-called Gen2). Right 
before every exposure, the ICS will execute modules of the ETS to update target assignment 
to match with the configuration of exposure such as flexure changes by a telescope pointing. 
Once exposure finished, acquired image files will be supplied to an on-site version of the 
DRP package immediately, and will be reduced in parallel to the next exposure but only on 
selected area of images for quick quality assurance, which will be provided to observers and 
the SPT to update following exposures. Also, full reduction works by the DRP will be 
performed later at off-site, these output will be stored to database systems in the SPT 
for scientific uses.
In the survey by PFS, a number of targets both observed at once and assigned to one exposure 
is significantly large, capabilities for on-site automatic rearrangements and redesignings, 
such that switching scientific targets marked as succeeded to acquire required signal to 
noise ratio, are important for efficient survey operation. For these, PFS is planning to 
have somehow simpler version or module picked up for works of the SPT and the DRP packages.

\section{System Overview}

To perform on demand exposure reconfiguration, we need to have on-site DRP to be capable to 
output necessity metric values. Our baseline for on-site data reduction for quick quality 
assurance for each exposure is to finish within next exposure, assuming standard exposure 
time of 10 to 15 minutes, and to output rough estimation of noise spectrum and conditions 
of exposure, such as sky background conditions over an entire field or these for sky 
backgrounds including OH emission lines, but not a full reduction of acquired images. 

On observatory point of view, PFS is assumed to be operated under fully queue mode of the 
Subaru observatory, it is also important to be capable of presenting metrics of exposure, 
such as average seeings over exposure, sky conditions, or estimated noise. On-site data 
reduction is required to output these information, as well as items we need for a survey 
operation. 
Also, as a facility instrument, PFS instrument operation at the telescope and also 
preparation of exposures need to be performed without such intensive information from 
survey designs, and also to be operatable at least for core functionalities, such as an 
exposure execution or a data reduction, independent from an integrated database system.

\subsection{Database design}

Designs of the PFS surveys are still under development, and also in a phase of collecting 
requirement from a scientific survey design, such how we will dispatch targets into survey 
coordination or how we will do and reflect quality assurance on acquired images, we have 
started designing our survey database with two key design principles: schema design to be 
capable of accepting further additional information or requirement on preparations or 
reductions of objects such as broad or medium resolution band magnitudes from imaging data 
for a signal to noise ratio measurement, and subset of the entire database design to be 
used as a database design for instrument operation on site. 

\begin{figure}
\begin{center}
\includegraphics{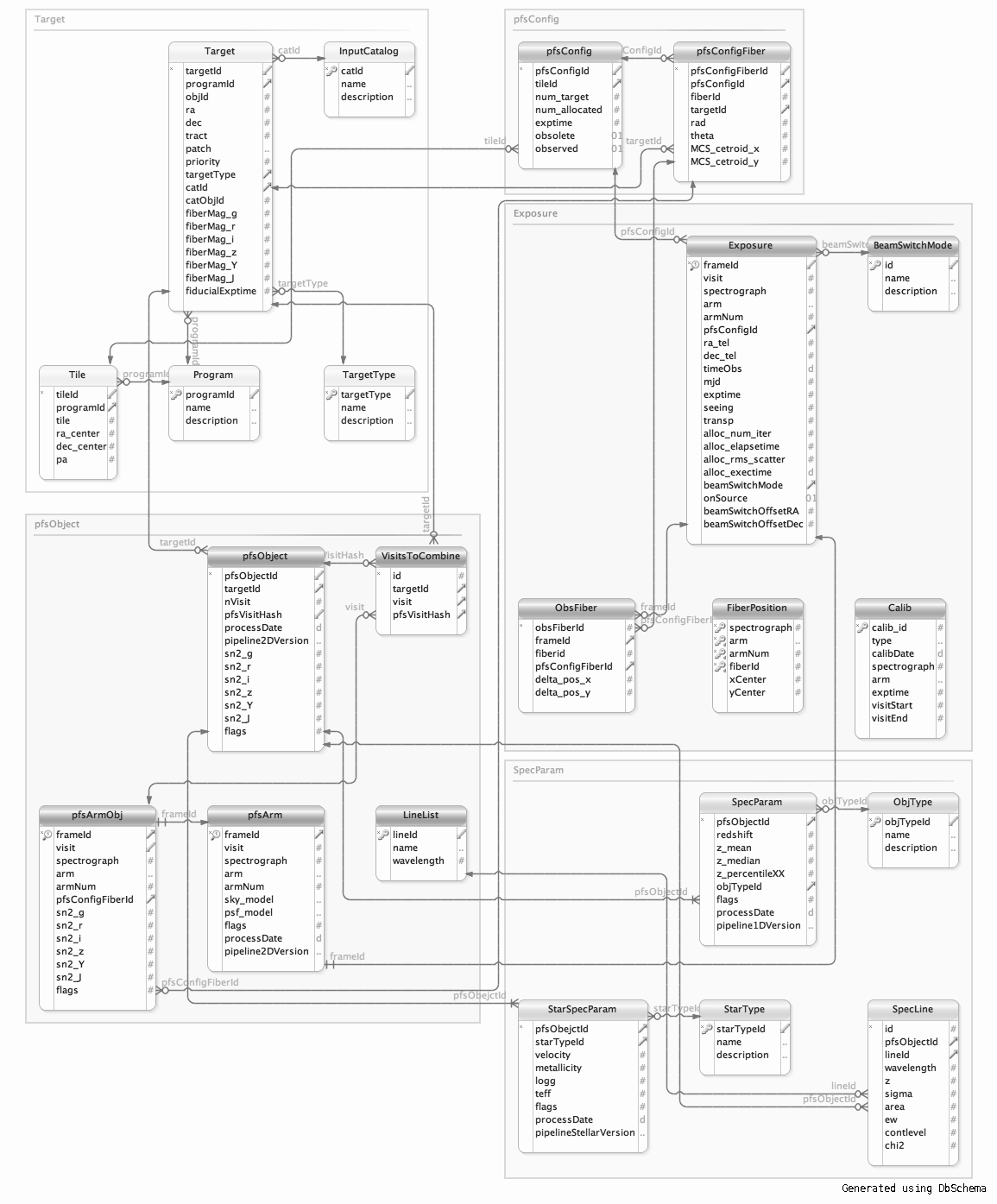}
% use 200dpi, in PNG header
\end{center}
\caption
{\label{fig:db_schema}
Schema definition of the entire PFS database system. 
Each small box with its header is a definition of a table where items are columns and 
box header is a name of table, and covering boxes are groupings of each table grouped by 
functionalities. Relationships between tables are drawn as arrows. 
}
\end{figure}

Current design is shown in Figure.~\ref{fig:db_schema}, and each group with a covering box 
shows a group of a data category, like list of target objects ('Target'), an exposure 
configuration ('pfsConfig'), an actual exposure information ('Exposure'), output from 
the data reduction pipeline ('SpecParam'), and tracking per object for the entire survey 
progress tracking and preparation ('pfsObject'). 
As this version is a preliminary version, no normalization nor optimization is applied 
to details of table shown in the figure. We will develop by adding keys and tables 
required from survey target allocation and data reduction point of view, such as metric 
values for target selection or definition of metric values for data quality assurance, 
as well as by fixing availability or operability with building and running a prototype 
simulator for each data processing.

For on-site exposure operation, the ICS of PFS need to be operated independed from the 
database system for the SSP survey to run individual observation programs or to run in 
a queue mode operation, an independent database system with several table definitions 
in boxes of Figure.~\ref{fig:db_schema} related on observation execution procedures will 
be hosted in the ICS system for operation, such as 'pfsConfig' for an input and 'exposure' 
for an output per each exposure and its tracking. 
Although, how we will coordinate flow of these information from the survey coordination 
system or its database and an observation preparation system for individual observation 
programs to the on-site database system is not well defined yet, two options are 
considered: file based operation copied to the ICS at the telescope from external or Hilo, 
or database replication between one at the telecope and one at Hilo with some external 
interface to get inputs such as web based API. 
Ouput data to observers from the ICS for each exposure shall be files in FITS format as 
defined by official archive, but entire telemetries or statuses are not possible to be 
included, such as raw seeing metrics measured by auto guide camera per evrey one to several 
seconds or periodical logging of instrument status. Such telemetry values will be provided 
through the Subaru telemetry system (STS), but could be copied directly from an on-site 
database to the survey database or could be provided in a certain format exported from 
the on-site database if necessary. 
Also, the output data files will be input to the DRP both on-site and off-site, a format 
of these data files is under development from both instrument operation and data reduction 
point of view. Set of minimum requirements as the telescope operation and archive point 
of view is well devined, but such additional information required by the DRP will be 
added into the datamodel, which is an internal model registration for file formats 
exchanged between PFS software packages. 
The on-site DRP is assumed to be processed at Hilo but not at the telescope, following 
data transfer from the telescope to Hilo facility by the observation operation system 
(Gen2) of the Subaru telescope, and a way to present output from the on-site DRP to 
observers is also under discussion.

\section{Instrument Control Flow}

Development activities of hardware of the PFS instrument are performed at many collaborating 
institutes in physically distributed all over the world, such as LAM (Marseilles, France), 
Princeton University (Princeton, NJ, U.S.A.), JHU (Baltimore, MD, U.S.A) 
and LNA (Itajuba, MD, Brazil) for the spectrograph system, or Caltech and JPL (Pasadena, 
California, U.S.A.), ASIAA (Taipei, Taiwan), and LNA for the PFI. The entire PFS hardware is 
consisted mainly with three subsystems, which are spectrograph, PFI, and metrology camera 
system, and will be connected both optically through fibers and by control system first time 
at the Subaru telescope. Before that, each subsystem is planned to be assembled, integrated, 
and tested at an institute responsible for subsystem integration, and also we need to verify 
control software of the subsystem as much as possible to prevent back and forth between there 
and the Subaru telescope. 

As progress of hardware development and integration, functionalities of each hardware component 
is partly implemented and under testing. From implementation and testing on hardware, we found 
additional limitations by hardware or required coordinations, and updates on instrument control 
sequences and handling of coordinates are under working.

\subsection{Instrument operation sequence}

\begin{figure}
\begin{center}
\includegraphics{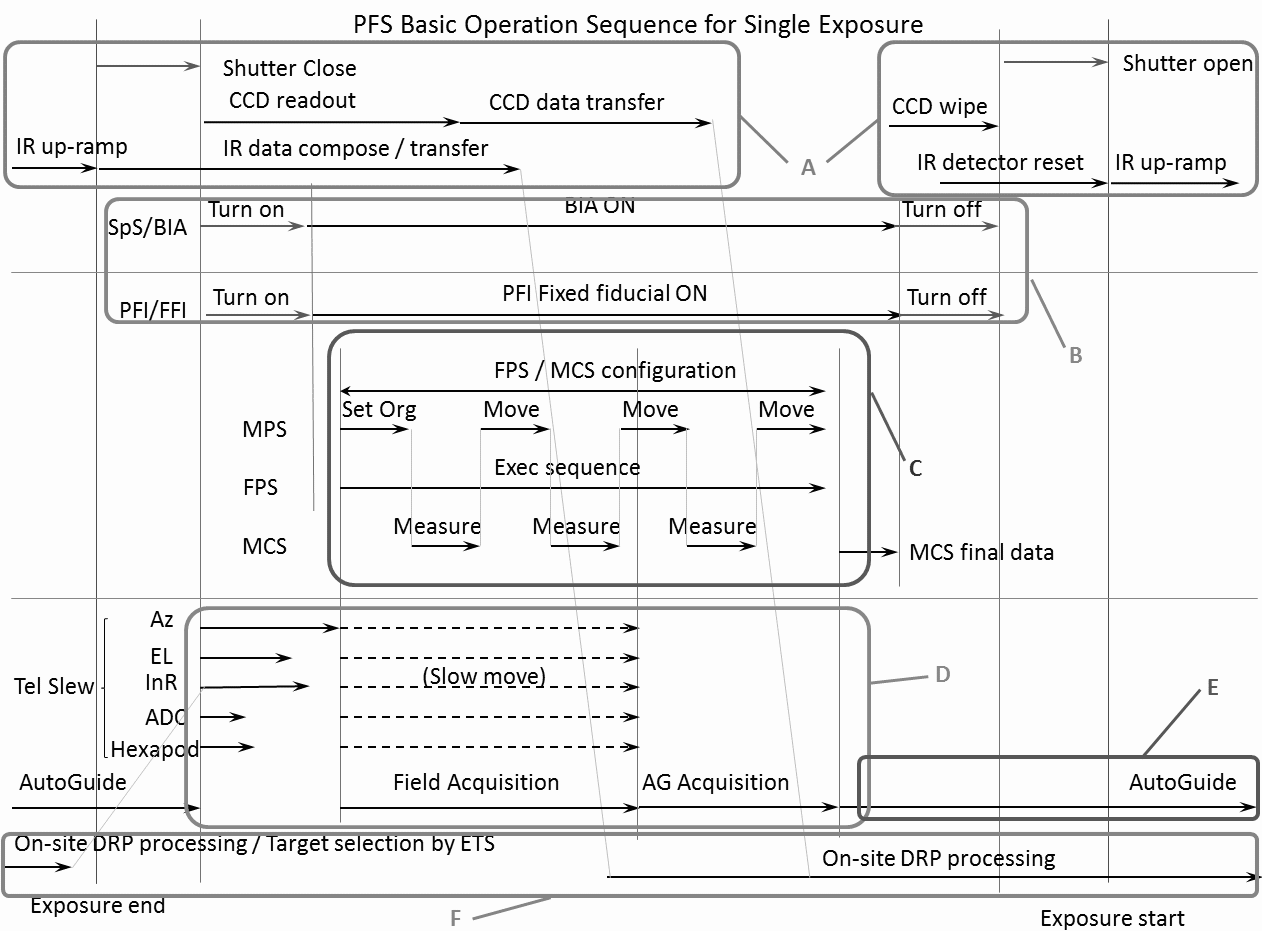}
% use 200dpi, in PNG header
\end{center}
\caption
{\label{fig:ics_sequence}
Schematic operational sequence of one exposure. 
Each rounded square shows one sub-sequence whose operation is closed within one hardware 
subsystem which will be integrated at one institute right before delivery to Subaru. 
}
\end{figure}

An entire instrument operation sequence between two exposures, from the end of last exposure 
(closing shutters) to the beginning of next exposure (opening shutters), are summarized as 
Figure.~\ref{fig:ics_sequence}. 

Cobra positioners of the PFS have no position encoder within two rotational piezo motors, 
external calibration of current positions is required, and the PFS takes an optical measurement 
using the MCS from the Cassegrain focus by getting images of back illuminated fiber spots. 
In parallel to detector readout processes and their output as FITS files with operating 
three arm camera systems and mechanical elements in the spectrograph system, 
Cobra positioning sequence runs with operating the PFI and the MCS. 
Positions of target objects commanded to the sequence are updated during this process, by 
calling a coordinate conversion module in the ETS package with statuses such as a telescope 
pointing and an instrument rotator of the PFI at the time, for applying small changes from 
flexure or distrotion changes. 
Slewing telescope to acquire field is performed in parallel or serial to these sequences, 
depends on its affect to images taken by the MCS, using guide objects on six Auto Guider 
Cameras (AGC) mounted on the PFI focal plane. 
Finally, when all subsequences are finished, a next exposure starts with operating shutters 
and camera systems.

Final integration and validation of these operation is possible only at the telescope, 
that controlled hardware subsystems will be connected optically first time at the telescope. 
To make these works as smooth and short as possible, we divided the entire operational 
sequence of one exposure into six subsequences, marked as round squared box in 
Figure.~\ref{fig:ics_sequence}, with simple interlock in between and no overlapped 
hardware component. Also each subsequence is defined to be possible to validate at 
each subsystem integration place, except for two: field acquisition and auto guiding, 
and Cobra configuration, which require external system (telescope) or two subsystems. 
Even for them, we plan to validate their functionality by hardware emulation during each 
subsystem integration, such that we plan to have simple optical system, by a mirror and 
a set of field lens but not like the WFC, to feed spots from back illuminated fibers into 
the same CMOS sensor used for the MCS to test Cobra positioning sequence.

\subsection{Coordinates and their transformation in an operation}

\begin{figure}
\begin{center}
\begin{minipage}{.45\textwidth}
\includegraphics{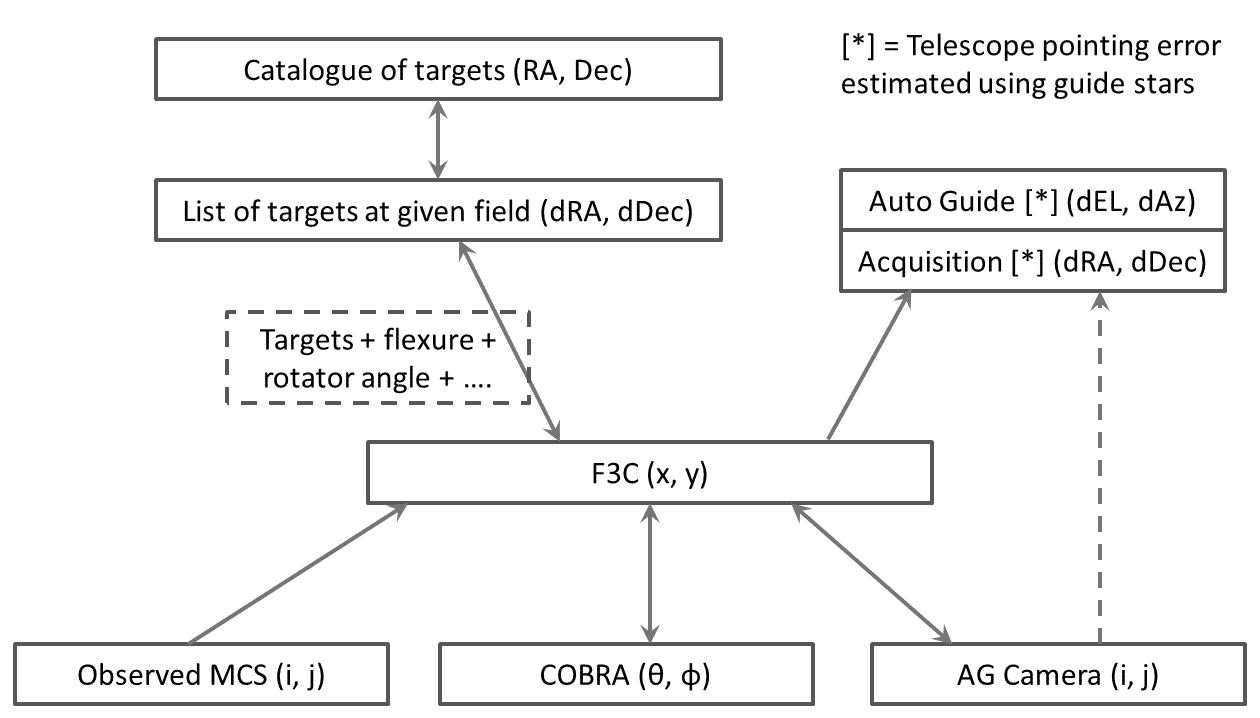}
\end{minipage}
\begin{minipage}{.45\textwidth}
\includegraphics{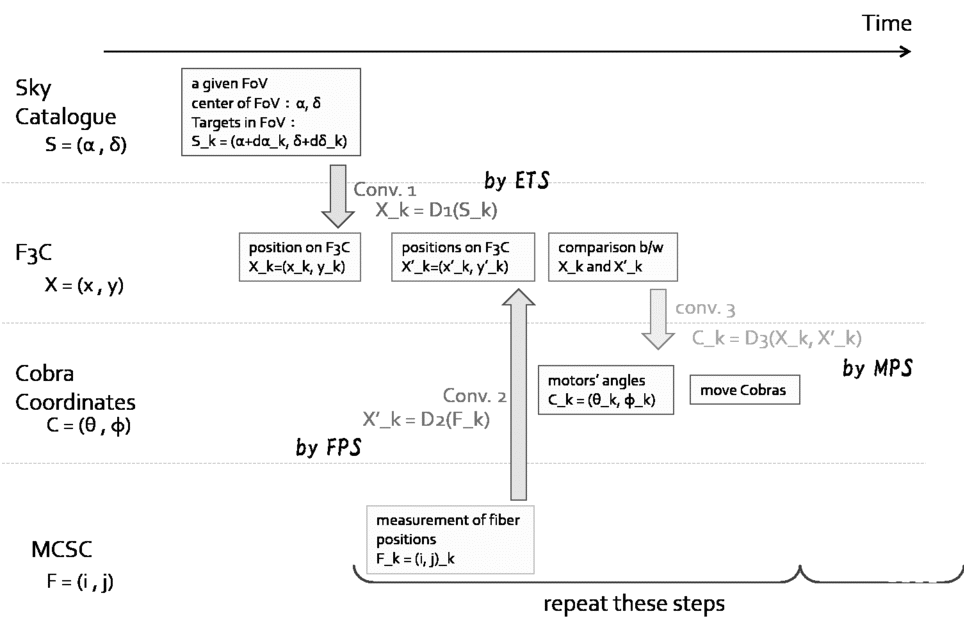}
\end{minipage}
% use 300dpi, in PNG header
\end{center}
\caption
{\label{fig:pfs_coords}
(Left) 
Relations among coordinates used for PFS instrument operation. In each box, a value for 
a coordinate shown as $(i, j)$ is for their detector pixel value, and as $(x, y)$ is for 
detector independent one. 
(Right) 
Conversions among coordinates used during an instrument operation for Cobra positioner 
"move to target" operation. 
}
\end{figure}

Four coordinates are used for fiber positioning by Cobra positioners in the operation: 
coordinates of catalogues for science target objects, the Fixed Fiducial Fiber Coordinate 
(F3C), Cobra coordinates for each Cobra with two angles of rotation motor, and the 
Metrology Camera System coordinate (MCSC). Relations among coordinates are shown in 
Figure.~\ref{fig:pfs_coords}. 

% XXXXXXX
\newcommand{\bm}[1]{\text{\boldmath{$#1$}}}
\newcommand{\degree}{\ensuremath{{}^o}}

Cobra positioners are mounted on the optical bench of the PFI, which is placed at the 
focal plane, although conversion between the focal plane and the Cobra coordinate 
$(\theta, \phi)$ is static and unique wihtout any effect by flexures of the PFS 
instrument or the telescope, conversions to the focal plane from the sky or the MCSC 
will change by environmental status including a pointing of the telescope. Between 
these coordinates, an intermediate coordinate is required for ease of operation, such 
as the sky coordinate or the MCSC coordinate with one specified environmental condition, 
or a physical coordinate fixed to some hardware. Still trade off studies are under going, 
the F3C is assumed to be used as the intermediate coordinate for conversion and exposure 
configurations. 
Cobra fiber positioners is operated with the Cobra coordinate, which is defined as the 
target values (angles) of two rotational motor: $(\theta, \phi)$. The motion of each 
Cobra is controlled by the Movement Planning Software (MPS; developped by JPL/Caltech) 
with the two angles of stages in degree. 
To convert target positions on the intermediate coordinate to the Cobra coordinate, 
conversions from a normalized orthogonal coordinate to two rotational axis coordinate 
is simple and easy, but conversion from non normalized nor orthogonal coordinate require 
us to convert through some normalized orthgonal coordinate or use local conversion 
parameter with ignoring small error from non-linearity. 
The MCS subsystem mounted at the Cassegrain focus will lock the focal plane through 
the wide field corrector (WFC), and calculates centroids of back-illuminated fibers 
for both fixed fiducial fibers and science fibers. The centroids are determined as the 
position on a CMOS sensor in the unit of pixels, equivalent to the MCSC. The centroids 
of the fibers, therefore, is determined in MCSC as $\bm{F_k}=(i_k [ \mathrm{pixel} ], 
j_k [ \mathrm{pixel} ])$, which affected by flexure of hardware. 

In the F3C coordinate, target objects and their fiber positions are described in the 
unit of mm: $\bm{X_k}=(x_k [ \mathrm{mm} ], y_k [ \mathrm{mm} ])$, which is defined as 
mean plane of the top surface of 97 fixed fiducial fibers, observed by the MCS and 
converted from those in the MCSC. 
F3C resembles physical metrics of the PFI, although positions of the fixed fiducial 
fibers of the PFI will be roughly measured at off-site (ASIAA) during integration by 
accuracy of ~10$[\mu m]$, we might need more acculate physical fixed points and will 
be updated using measured relative relationship of fixed fiducial fibers by the MCS 
from one measured at an off-site. Parameters for Cobra operations, such as center 
positions of stages of Cobras, will also be measured on this F3C coordinate. 

Figure.~\ref{fig:pfs_coords} right shows observational sequence in particular during 
positioning fibers, and transformations of the above coordinates. The procedure of fiber 
positioning is as follows:

\begin{enumerate}
%%% 1. sky
\item The ETS calculate the target positions $\bm{S_k}=(\Delta \alpha _k [ \degree ], 
  \Delta \delta _k [ \degree ])$ in the observed field centered at $\bm{S}=(\alpha 
  [ \degree ], \delta [ \degree ])$.
%%% 2. sky -> F3C
\item The ETS transforms the target positions in catalogues to those in the F3C using 
  the telescope parameters at that moment ($Az, El, InR, T$). We define this transform 
  as $D_1$. That is,
  \begin{equation}
  \begin{array}{crcl}
  & \bm{S_k}=(\alpha _k [ \degree ], \delta _k [ \degree ]) & \rightarrow & \bm{X_k}=(x_k [ \mathrm{mm} ], y_k [ \mathrm{mm} ]), \\
  & & D_1 & \\
  \mathrm{or} & \bm{X_k} &= & D_1(\bm{S_k}).
  \end{array}
  \end{equation}
%%% 3, MCSC -> F3C
\item\label{item:mcs2f3c} Take fiber image with the MCS and measure fiber positions 
  $\bm{F_k}=(i_k [ \mathrm{pixel} ], j_k [ \mathrm{pixel} ])$. Then FPS (Fiber Positioning 
  Sequencer) transforms the fiber positions in MCSC to those in F3C. We define this 
  transform as $D_2$. That is,
  \begin{equation}
  \begin{array}{crcl}
  & \bm{F_k}=(i_k [ \mathrm{pixel} ], j_k [ \mathrm{pixel} ]) & \rightarrow & \bm{X'_k}=(x'_k [ \mathrm{mm} ], y'_k [ \mathrm{mm} ]), \\
  & & D_2 & \\
  \mathrm{or} & \bm{X'_k} & = & D_2(\bm{F_k}).
  \end{array}
  \end{equation}
%%% 4. F3C -> MPS
\item\label{item:f3c2cobra} The MPS calculates Cobra Coordinates of fibers $\bm{X'_k}$ 
  and targets $\bm{X_k}$. We call this transformation $D_3$.
  \begin{equation}
  \begin{array}{crcl}
  & \bm{X_k}=(x_k [ \mathrm{mm} ], y_k [ \mathrm{mm} ]) & \rightarrow & \bm{C_k}=(\theta _k [ \degree ], \phi _k [ \degree ]) ,\\
  & & D_3 & \\
  \mathrm{or} & \bm{C_k} & = & D_3(\bm{X_k}).
  \end{array}
  \end{equation}
  Comparing these positions, The MPS derives the angles to move cobras $\bm{C_k}$:
  \begin{equation}
  \begin{array}{crcl}
   & \bm{C_k} & = & D_3(\bm{X'_k}) - D_3(\bm{X_k}).
  \end{array}
  \end{equation}
  Then move Cobras following the derived angles $\bm{C_k}$.
%%% 5. repeat
\item Repeat \ref{item:mcs2f3c} and \ref{item:f3c2cobra} until the differential between the 
  target and the fiber positions meets the required accuracy; $\Delta r_k= \sqrt{ \Delta x_k^2 
  + \Delta y_k^2} \leq 10 \;\mathrm{[um]}$. The FPS judges which fibers should be moved, and 
  sends $\bm{X_k}$ and $\bm{X'_k}$ to theMPS.
\end{enumerate} 

During the commissioning, the transformation functions $D_1$, $D_2$, and $D_3$ are defined 
and/or calibrated.

\appendix    %>>>> this command starts appendixes

\section*{ACKNOWLEDGEMENTS}

We gratefully acknowledge support from the Funding Program for World-Leading Innovative 
R\&D on Science and Technology (FIRST) program "Subaru Measurements of Images and Redshifts 
(SuMIRe)", CSTP, Japan.
The work in ASIAA, Taiwan is supported by the Academia Sinica of Taiwan.
This work was supported by JSPS KAKENHI Grant Numbers 15H05893, 15K21733, 15H05892.
Kavli IPMU is supported by World Premier International Research Center Initiative (WPI), MEXT, Japan.

% References
%\bibliography{report} % bibliography data in report.bib
\bibliography{spie-2016-pfs-soft}
\bibliographystyle{spiebib} % makes bibtex use spiebib.bst

\end{document}